\documentclass[twoside]{LCWS11}
\usepackage[latin1]{inputenc}
\usepackage[dvips]{graphicx,epsfig,color}
\usepackage{wrapfig,rotating}
\usepackage{amssymb,amsmath,array}
\usepackage{cite}
\usepackage[tight,hang]{subfigure}
\usepackage{subfigure}

\makeatletter

\newcommand{\tblcaption}[1]{\def\captype{table}\caption{#1}}
\newcommand{\figcaption}[1]{\def\captype{figure}\caption{#1}}
\makeatother

\begin{document}
\title{
Little Higgs with T-parity measurements at the ILC} 
\author{Eriko Kato$^1$  Masaki Asano$^2$  Keisuke Fujii $^3$Shigeki Matsumoto$^4$  Yosuke Takubo$^3$ and Hitoshi Yamamoto$^1$
\vspace{.3cm}\\
1- Tohoku University -Department of Physics ,Sendai 980-8578- Japan\\
2-II. Institute for Theoretical Physics, University of Hamburg, \\
Luruper Chausse 149, DE-22761 Hamburg,-Germany\\
3-High Energy Accelerator Research Organization (KEK),Tsukuba 305-0801- Japan\\
4-IPMU, TODIAS, University of Tokyo, Kashiwa, 277-8583- Japan\\
}

\maketitle

\begin{abstract}
The Littlest Higgs with T-parity Model (LHT) is one of the attractive candidates of physics beyond the Standard Model.
 In this study we focus on  heavy gauge bosons and heavy leptons that LHT imposes, which are detectable at ILC.
 The mass and couplings of these particles possess important information of the model's mass generation mechanism and how they interact with other particles.
 Precision measurement will be essential among verifying the model. We will address the measurement accuracy of Little Higgs particle masses, couplings and model parameters at ILC. 
\end{abstract}

\section{Introduction}

 The Standard Model has made remarkable success in accurately describing particle physics. However the mass of the Higgs boson suffers from an instability under radiative corrections. This "hierarchy problem" leads us to expect new physics near the weak scale. In the Little Higgs scenario, the Higgs boson is regarded as a pseudo Nambu-Goldstone boson arising from the spontaneous breaking of a global symmetry. Due to a imposed symmetry, new particles such as heavy gauge bosons and top-partners are introduced.
 Contributions of these particles cancel out the main quadratically divergent corrections of the Higgs boson mass at a one-loop level. Thus solving the hierarchy problem. 
 
 In this study we discuss the Littlest Higgs model
 and furthermore implement a $Z_2$ symmetry called T-parity, which was imposed to satisfy constraints from electroweak precision measurements \cite{cheng2003, cheng2004_1, cheng2004_2}. The lightest T-odd particle(heavy photon) is stable and provides a good candidate for dark matter. In order to test the Little Higgs with T-Parity (LHT) model, precise determinations of properties of little Higgs partners are essential, because these particles hold information of mass generation mechanism and thus are directly related to the cancellation of quadratically divergent corrections to the Higgs mass term. The measurements of heavy gauge boson masses and heavy leptons are particularly important. Heavy gauge bosons acquire mass through the breaking of the global symmetry and can solely be described with the parameter $f$,  i.e. the energy scale of the global symmetry breaking. Heavy leptons also acquire mass terms through the breaking of the global symmetry and are described with $f$ and the lepton yukawa coupling($ \kappa $). 
 All the masses and interactions in the gauge boson sector and lepton sector are described with these two parameters. We can extract the two mass originating parameters by measuring LHT masses. Furthermore LHT particles have characteristic interactions similar to the standard model, and they can also be explained with these two parameters.
 
 Here we will discuss the precision measurement of three heavy gauge bosons and two heavy leptons. Checking the consistency with extracted parameters from measured masses and couplings will be strong evidence that the discovered new particles are indeed LHT particles.
 
At the Large Hadron Collider (LHC), although it has potential of discovering LHT particles, because there are missing particles in the final state, the center mass energy is unknown and there are large background events, it is extremely difficult to reconstruct the masses of T-parity odd LHT particles. 
 On the other hand, the ILC with it's clean environment and fixed center mass energy, will be an ideal environment to measure the properties of the LHT particles. \\
 In this study we will address ILC's measurement sensitivity to masses, parameters, couplings of heavy gauge bosons and heavy leptons based on realistic Monte Carlo simulation.

\section{Representative point and target mode}

 In order to perform a numerical simulation at ILC, we need to choose a representative point in the parameter space of the LHT. Firstly, the model parameters should satisfy the current electroweak precision data. 
It also has to satisfy the cosmological observation of dark matter relics. Thus, we consider not only the electroweak precision measurements but also the WMAP observation to choose a point in the parameter space.
The current measurement of relic density is $\Omega_{DM}h^2 $= 1.06$\pm$0.008 from this measurement, we have selected a representative point for the Higgs mass and vacuum expectation value ($f$), where they are 134 GeV and 580 GeV, respectively. 
 
 As for the lepton Yukawa coupling($\kappa  $), we have selected a representative point of $ \kappa$=0.5,  
 If the yukawa coupling is too small heavy leptons become light and should be observed at previous experiments. If the yukawa coupling is too large, the four fermion interaction contribution increases and large discrepancy from the standard model emerges. 
 With the parameter sets of ($m_{H}$, $f$, $\kappa $)=(134GeV, 580GeV, 0.5), the masses of the heavy gauge bosons and leptons are ($M_{AH} , M_{WH} , M_{ZH} ,M_{eH}$,$M_{\nu H}$) = (81.9 GeV, 368 GeV, 369 GeV, 410GeV, 400GeV), where $A_{H}, Z_{H}, W_{H}, e_{H}, \nu_{H} $ are the Little Higgs partners of a photon, Z boson, W boson, electron, neutrino respectively. $A_{H}$ is a dark matter candidate. Since all the partners are lighter than 500 GeV, it is possible to generate them at ILC.\\

\begin{figure}[htbp]
\vspace{-2em}
	\begin{center}
	\includegraphics[width=0.72\columnwidth]{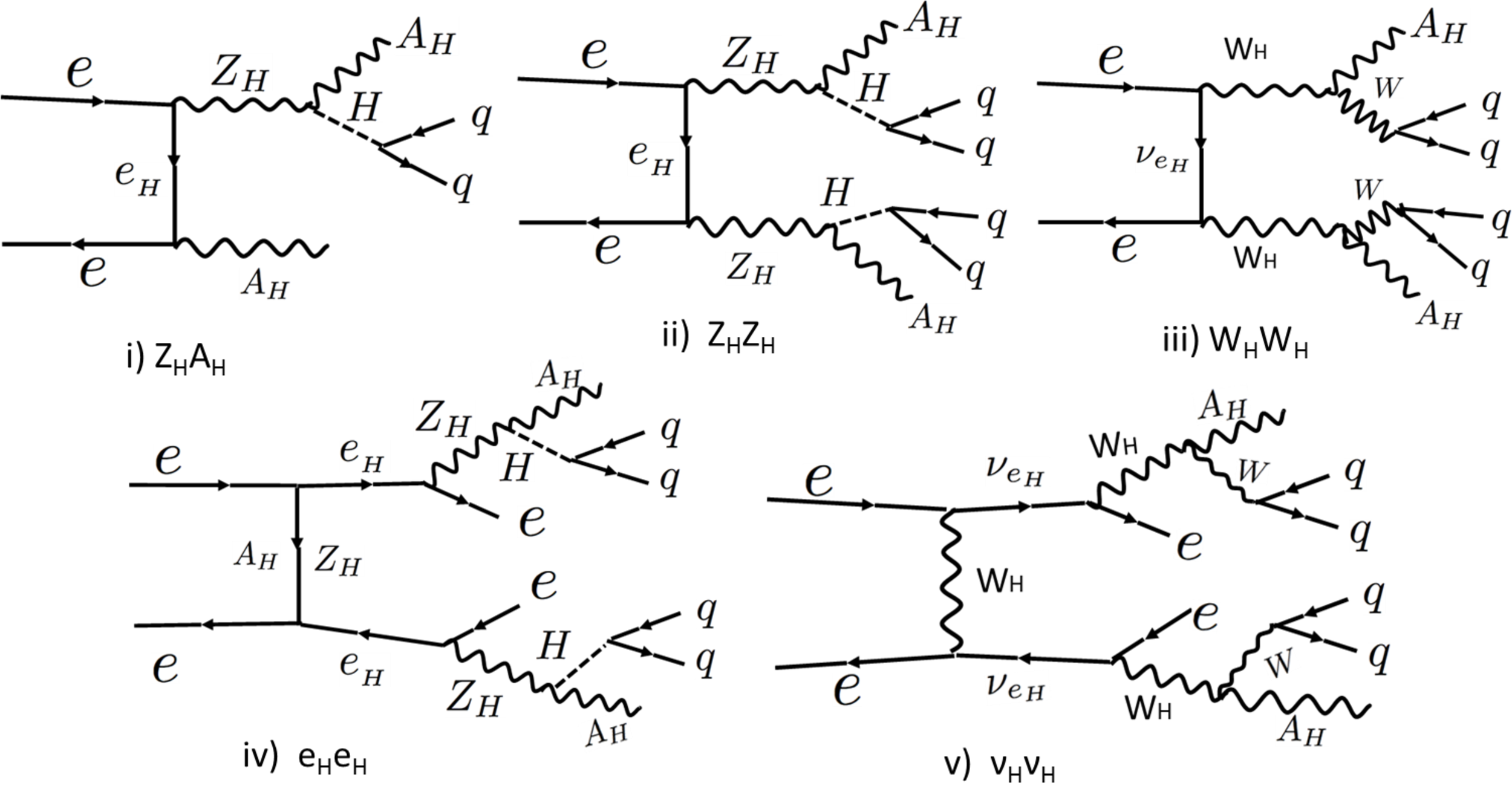}
	\figcaption{Feynman diagrams}
\end{center}\label{fig:diagram}
\end{figure}

To measure heavy gauge boson properties we used three processes; $e^{+}e^{-}\rightarrow A_{H}Z_{H} , Z_{H}Z_{H}$ and $W_{H}W_{H}$. $A_{H}A_{H} $ is also producible, but because $A_{H}$ is a dark matter candidate the signal is undetectable.
To measure heavy lepton properties we used two other processes; $e^{+}e^{-}$ $\rightarrow$ $e_{H}e_{H}$ and $\nu_{H}\nu_{H}$. 
$A_{H}Z_{H}$ was studied at $\sqrt{s}$ = 500 GeV and the other processes at  $\sqrt{s}$ = 1 TeV. 
The heavy gauge bosons $Z_{H}$ and $W_{H}100\% $branching ratios. $Z_{H}$  decays $100\%$ to $A_{H}$ and higgs, $W_{H}$ decays $100\%$ to $A_{H}$ and W.
Thus we have chosen the following final states for heavy gauge bosons.\\
i) $A_{H}Z_{H}\rightarrow A_{H}A_{H}h \rightarrow A_{H}A_{H}bb$  \\
ii) $Z_{H}Z_{H}\rightarrow A_{H}A_{H}hh \rightarrow A_{H}A_{H}qqqq$ \\
iii) $W_{H}W_{H}\rightarrow A_{H}A_{H}WW\rightarrow A_{H}A_{H}qqqq$\\
At the representative point we have chosen, heavy Leptons are heavier than heavy gauge bosons, therefore they can decay emitting heavy gauge bosons.
Branching ratios of $e_{H}$ and $\nu_{H}$ are shown at Table \ref{tab:brs}.
\begin{table}[htbp!]
\caption{Branching ratio of heavy leptons }
\begin{center}
\begin{tabular}[h]{l | l | l |l | l| l}
Particle & Branch & Branching ratio   & Particle & Branch & Branching ratio \\\hline \hline
  & $ eA_{H}$  & $30\%$  &  & $ e W_{H}$ & $65\%$\\\cline{2-3} \cline{5-6}
  $e_{H}$  & $ eZ_{H}$  &$25\%$ & $\nu_{H}$ & $ \nu Z_{H}$ & $11\%$\\\cline{2-3} \cline{5-6}
 & $ eW_{H}$  &$45\%$ & & $ \nu A_{H}$ & $ 24\%$\\
\end{tabular}\label{tab:brs}
\end{center}
\end{table}

For the heavy leptons we have chosen the following final states.\\
 iv) $e_{H}e_{H}\longrightarrow eeZ_{H}Z_{H}\longrightarrow eeA_{H}A_{H}qqqq $\\
 v) $\nu_{H}\nu_{H}\longrightarrow eeW_{H}W_{H}\longrightarrow eeA_{H}A_{H}WW$\\
Among the three possible ways a heavy electron can decay, the $\nu W_{H}$ branch has difficulty accessing the mass of $e_{H}$ since $e_{H}$ directly emits a missing neutrino and a darkmatter emitting $W_{H}$. 
The $eA_{H}$ branch suffers from large standard model and little higgs background. On the other hand, the $eZ_{H}$ branch has a very characteristic final state of two isolated energetic electrons and two higgs. In addition, the $eZ_{H}$ branch also has a large branch when the  lepton yukawa coupling is above 0.5, which is rather difficult to be probed at the LHC\cite{lhc_br}. 
The summary of the Feynman diagrams for the signal processes are shown in Figure \ref{fig:diagram}.

\section{Simulation tools}
\label{sec:figures}
Signal and background events are generated using the physsim \cite{physsim} event generator,  based  on  the  full  helicity amplitudes including gauge boson decays, calculated using HELAS \cite{helas} and BASES \cite{bases},which properly takes into account the angular distributions of the decay products. The effects of initial state radiation, beamstrahlung and beam energy spread are also included. 
In this study we will use unpolarized beams. Once the final state particles are produced their four momentum are handed down to Pythia 6.4 \cite{pythia} for parton showering and hadronization. For final state tau leptons the decay is carried out by TAUOLA \cite{tauola}. The detector response is simulated using JSFQuickSimulator\cite{fastsim} ; a fast Monte-Carlo detector simulator which implements ILD geometry \cite{geometry} and other detector-performance related parameters. The assumed detector resolution is specified in Table \ref{tab:limits}.\\
\begin{table}[htbp]
	\begin{center}
	\caption[ILD detector resolution]
	{\small ILD detector resolution}
	\begin{tabular}{lll}
Detector  & Resolution & Coverage\\\hline\hline
Vetex detector  & $\sigma_{ip}=5.0\oplus (10.0/p \beta sin^{3/2}\theta)  \mu m$  & $|\cos\theta| \leq  0.93$\\\hline
Central drift chamber  & $\sigma_{P_{T}}/P_{T}^{2}= 5\times10^{-5}P_{T} (GeV/c)^{-1}$  &$|\cos\theta| \leq  0.98$\\\hline
ECAL &  $\sigma_{E}/E= 17\% /\sqrt{E}\oplus 1\%$  &$|\cos\theta| \leq  0.99$ \\\hline
HCAL  & $\sigma_{E}/E= 45\% /\sqrt{E}\oplus 2\%$ & $|\cos\theta| \leq  0.99$\\
\end{tabular}
	\label{tab:limits}
	\end{center}
\end{table}

\section{Analysis strategy}
Three physical quantities will be extracted in this study, i.e. model parameters, LHT particle masses and LHT couplings. 
Since the overall analysis procedure is fairly similar among all processes, for simplicity we will name the pair-produced LHT parent particle LHp, and it's LHT daughter particle LHd.
As mentioned before, model parameters can be extracted from the measured LHT particle masses \cite{lht_mass_eq}, which can be derived by solving kinematics.
Couplings can be extracted from cross section and angular distribution of LHp. 
Here we will discuss the analysis strategy on how to extract these physical quantities from observables. 

\subsection{Mass measurement and parameter extraction }
Due to T-parity conservation, T-parity odd particles are produced in pairs. Once T-Parity odd particles are produced(LHp), for the same reason, they decay into a standard model particle and a T-parity odd particle(LHd).
 In general, we can extract both LHT particle masses (LHp and LHd) by analyzing  the energy distribution of the reconstructed standard model particle(SMd). The edges, i.e. the upper and lower kinematic limits of the standard model energy ($E_{max}$ and $E_{min}$) are written in terms of LHT particle masses.
\begin{eqnarray}
E_{max} & = & \gamma_{_{LHp}}E_{_{SM}}^{*}+\beta_{_{LHp}}\gamma_{_{LHp}}p_{_{SM}}^{*} \\
E_{min}  & = & \gamma_{_{LHp}}E_{_{SM}}^{*}-\beta_{_{LHp}}\gamma_{_{LHp}}p_{_{SM}}^{*}
\end{eqnarray}

where $\gamma_{_{LHp}}(\beta_{_{LHp}})$ is the $\gamma(\beta)$ factor of the LHp in the lab frame, while $E_{_{SMd}}^{*} (p_{_{SMd}}^{*})$ is the energy (momentum)
of the standard model particle in the rest frame of LHp. Keep in mind that the standard model particle energy satisfies, $E_{_{SMd}}^{*}= (M_{_{LHp}}^{2} + M_{_{SMd}}^{2}  + M_{_{LHd}}^{2} )/(2M_{_{LHp}})$
With this method we can extract new particle masses without any assumption of the model. 
Because the LHT model is described with only two model parameters it gives a characteristic mass hierarchy.
Performing this model independent mass measurement will provide strong evidence that the discovered new particles are indeed LHT particles.

\subsection{Cross section and angular distribution of LHp production}
By assuming the vertex structure, i.e. the spin and the ratio of a right-hand, left-hand couplings, we can extract the couplings concerning heavy gauge bosons and heavy leptons through cross section measurement.
There are a total of eight heavy gauge boson/heavy lepton vertacies  concerning the five processes we are treating. 
Therefore in addition to cross section measurement we will use the angular distribution of LHp(differential cross section)  for three processes in order to extract all couplings. 

The procedure on how to extract the couplings are summarized in Figure \ref{tab:coupl_strat}.
The red circles represent the vertex that will be measured at that process mode and the blue circles represent the vertex that are measured in previous process modes.     
Through differential cross section measurement, we can extract information about the couplings contributing to the production of LHp, i.e. the coupling sign, whether the contribution is destructive or constructive, and the ratio between red circled couplings and the blue circled couplings.

\begin{table}[htbp]
	\begin{center}
	\caption[coupling measurement strategy]
	{\small coupling measurement strategy}
\hspace{-3em}
	\begin{tabular}{l | l | l | l }
Process mode                                                        & Branch                                                                     & derived vertex                                                        & observable   \\\hline\hline
\includegraphics[width=0.15\columnwidth ]{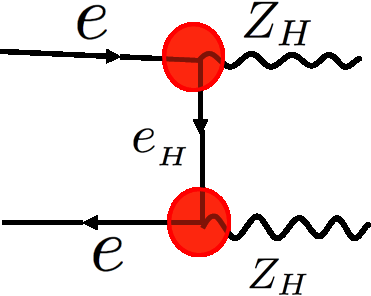}   &                                                                 & \includegraphics[width=0.15\columnwidth]{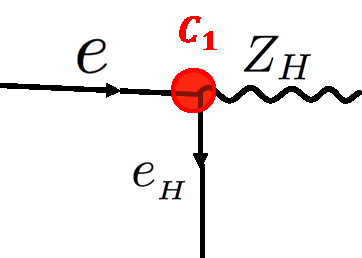}   & cross section    \\\hline
\includegraphics[width=0.15\columnwidth ]{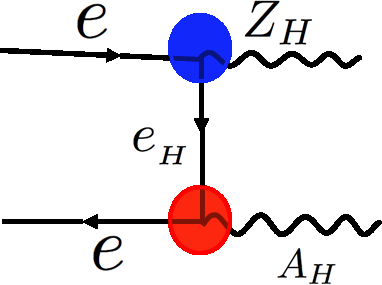}   &                                                                 &  \includegraphics[width=0.15\columnwidth]{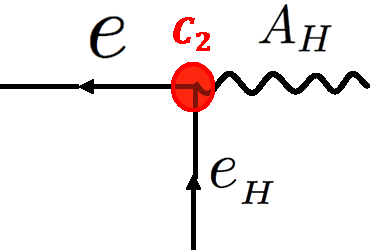}  & cross section    \\\hline
\includegraphics[width=0.35\columnwidth ]{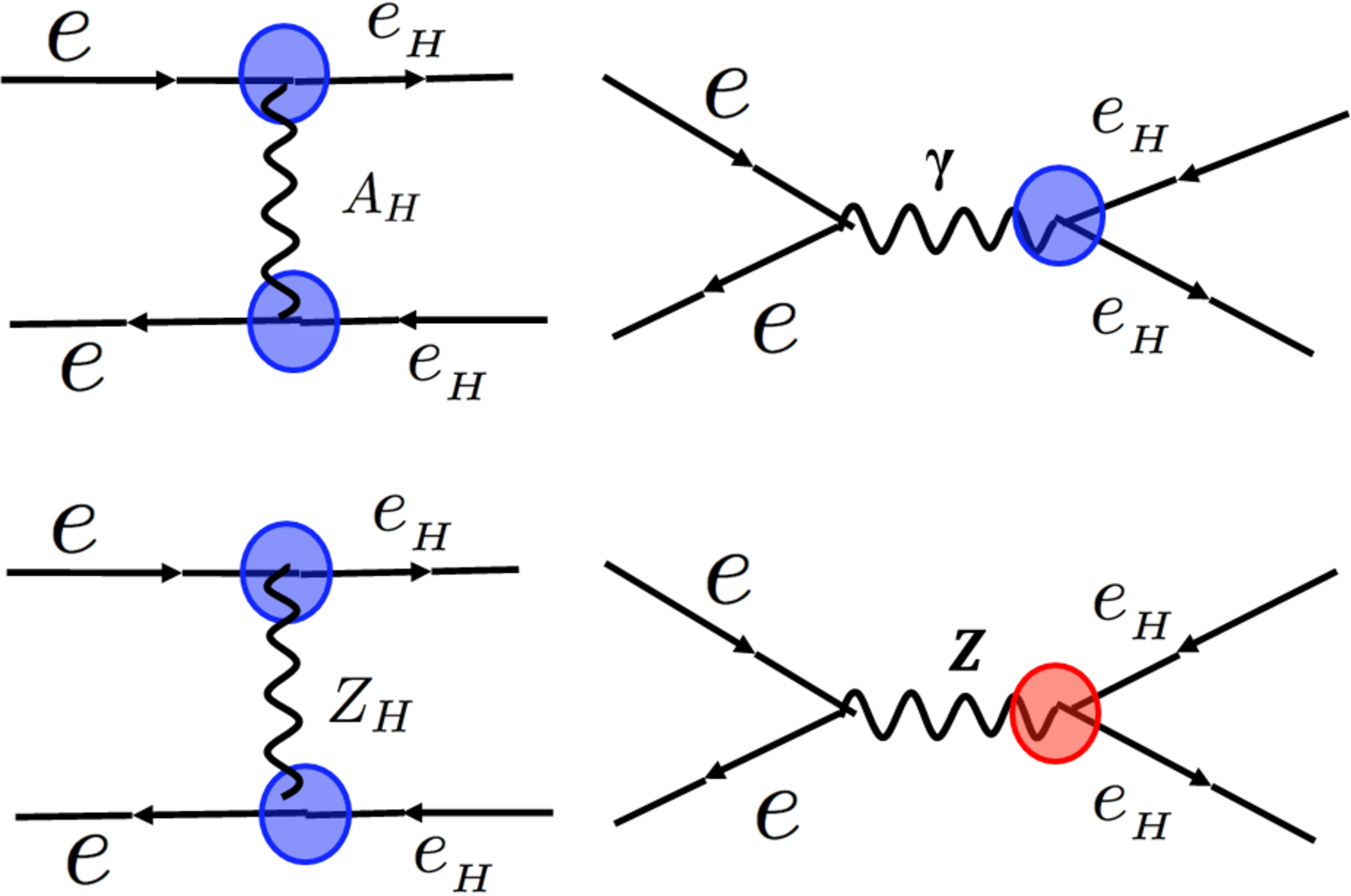}   & \includegraphics[width=0.15\columnwidth]{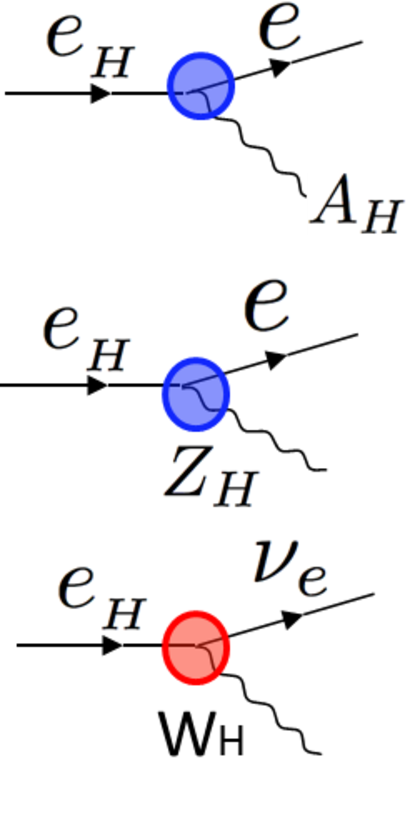}    & \includegraphics[width=0.15\columnwidth]{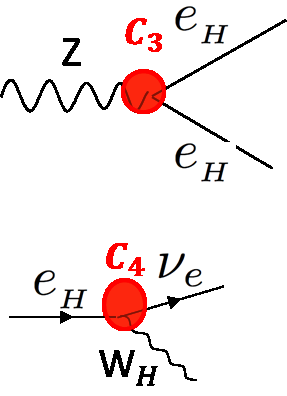}   & \shortstack{cross section, \\ differential cross section }\\\hline
\includegraphics[width=0.35\columnwidth ]{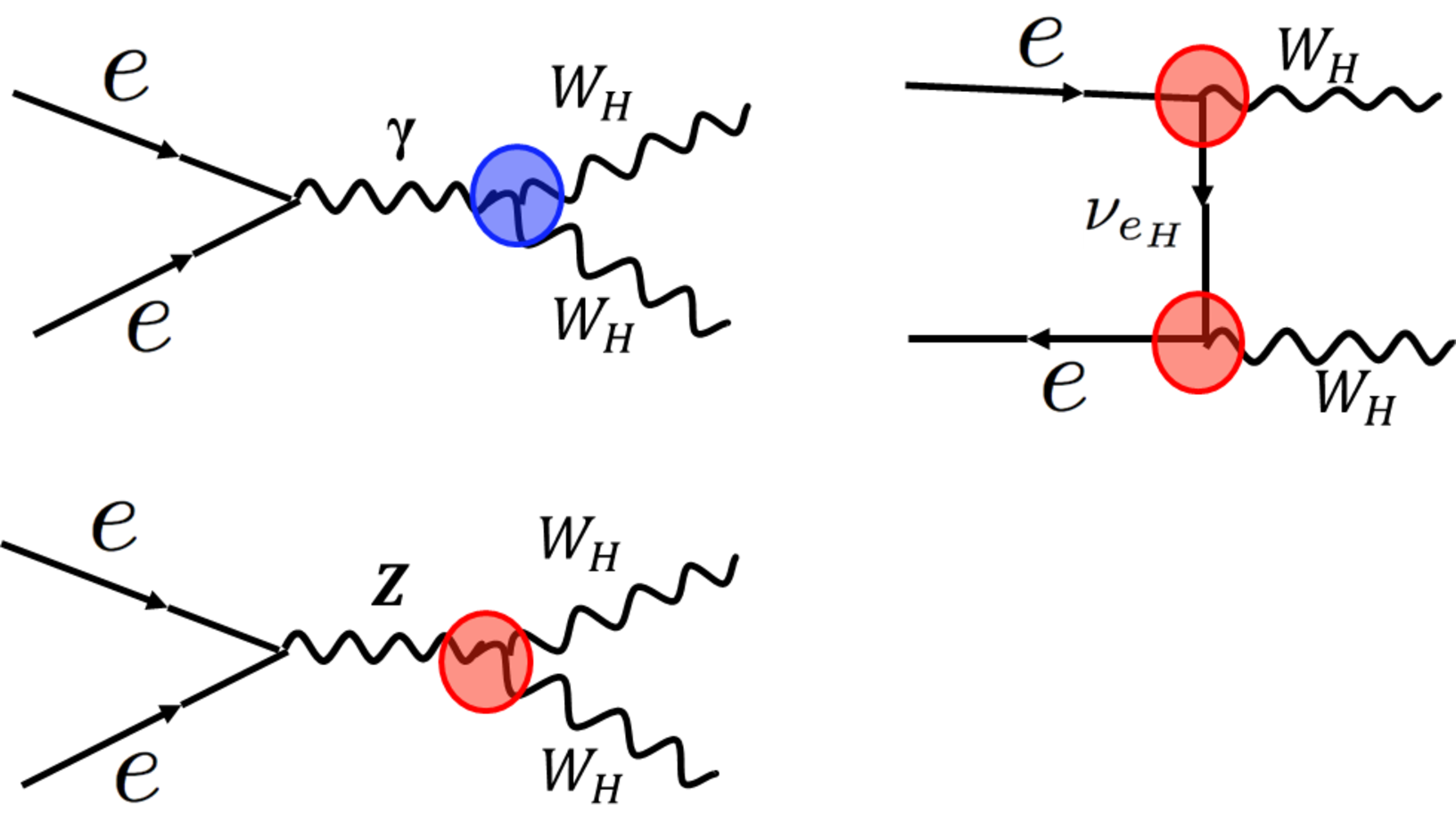}  &                                                                &  \includegraphics[width=0.15\columnwidth]{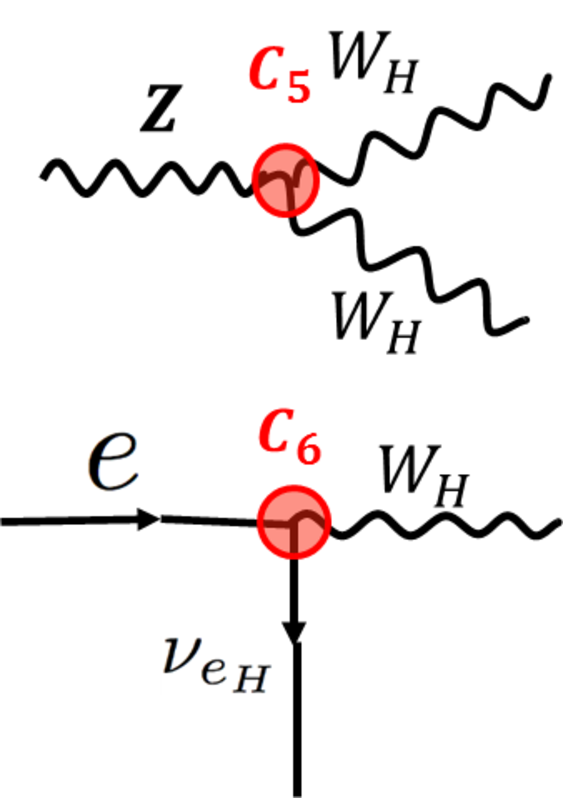}  & \shortstack{cross section, \\ differential cross section }  \\\hline
\includegraphics[width=0.40\columnwidth ]{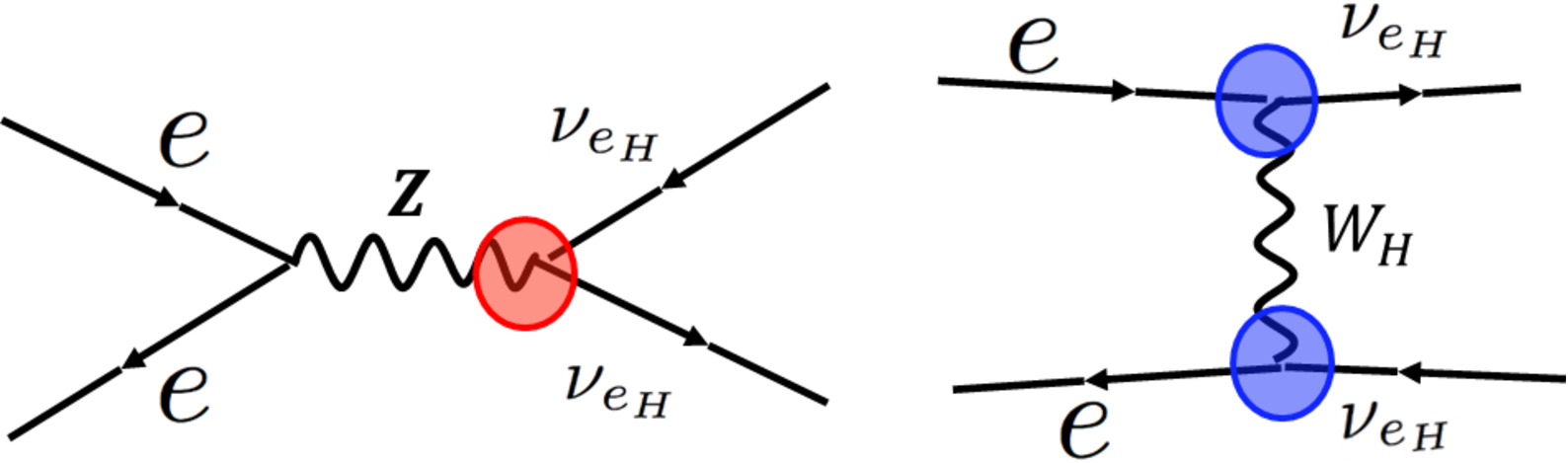} & \includegraphics[width=0.20\columnwidth]{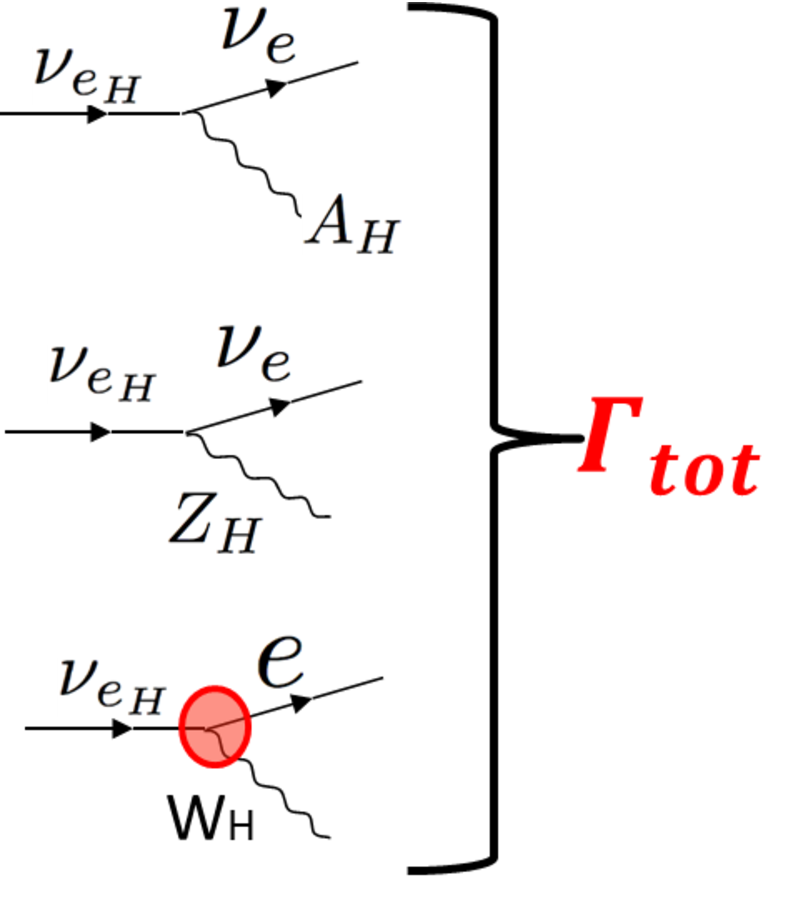} &  \includegraphics[width=0.15\columnwidth]{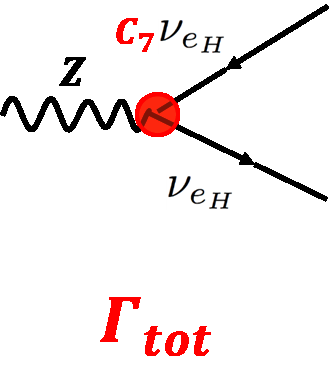} & \shortstack{cross section, \\ differential cross section }  \\
\end{tabular}
	\label{tab:coupl_strat}
	\end{center}
\end{table}

 First, by measuring the cross section of  $Z_{H}Z_{H}$ process mode, we can extract the coupling of $g_{ee_{H}Z_{H}}$.  
 
 Once we have measured this coupling, then we can use it as an input variable to extract the $g_{ee_{H}A_{H}}$ coupling in the  $Z_{H}A_{H}$ process mode. In the $e_{H}e_{H}$ process mode we use the previously  measured couplings $g_{ee_{H}Z_{H}}$ and $g_{ee_{H}A_{H}}$ as input variables to extract the couplings of  $g_{Ze_{H}e_{H}}$ and $g_{\nu e_{H}W_{H}}$. Through differential cross section measurement we can obtain the couplings' sign and the coupling ratio and  between $g_{Ze_{H}e_{H}}$  and other couplings that contribute to $e_{H}e_{H}$ production. Now that we have obtained all the couplings used in $e_{H}e_{H}$ production, we can extract  the couplings in the branch, i.e. $g_{e_{H}\nu W_{H}}$  through cross section measurement 
 
 Coupling extraction in $W_{H}W_{H}$ and $\nu_{H} \nu_{H}$ process modes are fairly the same. Except, for $W_{H}W_{H}$, cross section measurement is used for measuring the coupling contributing to the $W_{H}W_{H}$ production. 
 For  $\nu_{H} \nu_{H}$, the cross section measurement is used for the extraction of the total width of $\nu_{H}$.
 
 With this procedure, we are able to extract all couplings contributing to the five processes that we are treating.

 For simplicity, we will use the following $\chi ^2$ to evaluate the measurement accuracy of the couplings. 

\begin{eqnarray}\label{eq:chi2}
\chi^{2}=\Sigma_{i}\frac{(\vartheta_{(C_{meas})}^{i}-\vartheta_{(C_{th})}^{i})^2}{(\Delta \vartheta_{(C_{meas})}^{i})^2}
\end{eqnarray}

 Where $\vartheta$ is the observable, $C$ is the coupling, $meas$ is the subscript for a measured value, $th$ is the subscript for the true value, $i$ is the superscript that indicates the type of observable, e.g. cross section of $Z_{H}Z_{H}    $ etc. and $\Delta$ is used to represent the standard deviation of the measured observable.

Once we have extracted model parameters from masses and verified the consistency among all processes, we can say that it is extremely LHT like and make an estimation on the coupling values and search for the $\chi ^2$ minimum point around this estimation point. 
The correlation between two couplings will not be discussed in this study. It will be a subject once we have verified the consistency of couplings with model parameters independently.  

\newpage

\section{Results from simulation study}
 In this section, we present the results of our simulation study. 
Simulation has been preformed at $\sqrt{s}$=500~GeV for $A_{H}Z_{H}$ production and $\sqrt{s}$=1~TeV for other processes, each process is studied with an integrated luminocity of $500 ~fb^{-1}$.

\subsection{Signal selection}
The signal process we are treating include multiple jets in the final state. 
\begin{eqnarray}
y_{ij} =\frac{2E_{i}E_{j}(1-\cos \theta_{ij})}{E^2 _{vis}}   
\end{eqnarray}
The clusters in the calorimeters are combined to form a jet if the two clusters satisfy $y_{ij} \leq  y_{cut}$.
 $y_{ij} $ is defined as where $\theta _{ij}$ is the angle between momenta of two clusters, $E _{i(j)}$ are their energies, and
$E_{vis}$ is the total visible energy. 
For $A_{H}Z_{H}$ process, all events are forced to have two jets by adjusting $ y_{cut}$. For other processes, all events are forced to have four jets, which originated from a higgs or a W boson. In order to reconstruct the parent particle we have 
used a $\chi ^2$ function defined as
\begin{eqnarray}
\chi ^2 = (^{rec} M_{SM1} - ^{tr} M_{SM} )^2/\sigma ^2_{M_{SM}} + (^{rec} M_{SM2} - ^{tr} M_{SM} )^2/\sigma ^2_{M_{SM}} 
\end{eqnarray}
where SM is the parent particle, i.e. higgs or W boson, $^{rec} M_{SM1(2)}$ is the invariant mass of the first (second) 2-jet system paired as a parent candidate, $^{tr} M_{SM}$ is the true value of the parent particle mass, and $\sigma _{M_{SM}}$ is the mass resolution for the parent particle.

\subsection{Selection criteria}
Here we will discuss the selection criteria that is performed in each process mode. 
The selection criteria for $A_{H}Z_{H}$ and $W_{H}W_{H}$ process is the same as previous studies\cite{lht_mass_eq}.  

\begin{itemize}
\item[1.] $Z_{H}Z_{H}$ process
\end{itemize}

 The pair production of $Z_{H}$ is generated with a cross section of 98.0~fb at a center of mass energy of 1~TeV.
 Since $Z_{H}$ decays  into  $A_{H}$ and  a  higgs  boson,  the  signature  is  two higgs bosons in the final state. 
 At a mass with 134~GeV, the higgs boson mostly decays to  b-jets and  W bosons. Therefore in this analysis, we have used a 4-jet final state, which originates from the two higgs bosons.    
 We have selected events with $\chi ^2$ below 60 to select well reconstructed events. We also require signal events to have
 no isolated leptons, have more than 1 b-tag jet in the final state and have acoplanarity above 20 ${}^\circ $. 
The selection criteria  is summarized in Table \ref{tab:selection_zh}.

 The requirement of b-tagging is the existence  of  2  tracks  with  3 $\sigma$  displacement  from the  interaction point, where $\sigma$ is the impact
parameter  resolution. We select events which have more than one b-tagged jet.
 After b-tagging, the dominant background becomes $t\bar{t}$.  We investigated acoplanarity ($\pi -\phi$), where $\phi$ is the angle between two reconstructed higgs
candidates in the plane perpendicular to the beam axis. Since the signal has large missing momentum due to $A_{H}$, there is a large acoplanarity, compared to $t\bar{t}$.
The isolated-lepton rejection is preformed inorder to exclude events where the higgs decays into a W boson, followed by W's leptonic decay.
 These events have large missing energy, which causes the kinematical lower limit ($E_{min}$) in the energy distribution to shift to lower energy region. Causing degradation in the measurement accuracy of LHT particles masses.

 In this study, we identified isolated leptons events as leptons which satisfy the following equation. 
\begin{eqnarray}\label{eq:isolep}
Track Energy \leq 5+25/7 \times Cone Energy
\end{eqnarray}
Where TrackEnergy is the energy of the isolated lepton candidate and ConeEnergy is the energy around ($\cos \theta \geq  0.90$) the isolated lepton candidate.
An isolated lepton track has almost no energy surrounding it, as oppose to tracks from a jet where they are accompanied by other jet tracks.
 Equation (\ref{eq:isolep}) is optimized distinguish between leptons originating from $H \rightarrow WW^*$, where there is at least one W decays leptonicly, and leptons originating from  $H \rightarrow bb$.

\begin{table}[htbp]
	\begin{center}
	\caption[event selection of analyzed processes $Z_{H}Z_{H}$]{\small event selection of analyzed processes $Z_{H}Z_{H}$}
	\begin{tabular}{rccc}
  Process                                                                                        & cross section (fb)    &   $\sharp$ of events  & \shortstack{$\sharp$ of events\\ after all cuts}  \\\hline \hline  $Z_{H}Z_{H}\rightarrow A_{H}A_{H}HH$                                                   &   98.0                   &   49760                  &    18989                               \\
 $ \rightarrow A_{H}A_{H}qqqq$                                                            &                         &                        &   \\\hline
 $WWZ$                                                                                         &     63.86              &     31933          &  227 \\
 $\nu \nu WW$                                                                               &    14.67                &     7336            &  86 \\
 $WW$                                                                                           &    3069              &    1947408       &  994 \\
 $tt$                                                                                               &    192.9            &   96472         & 1749  \\
 $ZZ$                                                                                             & 202.2               & 101094            &  114 \\
  $ZH$                                                                                              &  17.98             & 8989              & 47 \\
 $W_{H}W_{H}$                                                                                  &  108.6             &  54343           & 87 \\
	\end{tabular}
	\label{tab:selection_zh}
	\end{center}
\end{table}

\begin{itemize}
\item[2.] $e_{H}e_{H}$ process
\end{itemize}

The signal we have used for $e_{H}$ pair production, is where both $e_{H}$ decay to $eZ_{H}$, which is further followed by $Z_{H} \rightarrow hA_{H}$.  The cross section of this process is 4.56~fb at a center of mass energy of 1~TeV.
Since there are two electrons and two higgs bosons in the final state, we have chosen 2 electrons and 4 jets as the signal. We furthermore require the two isolated electrons to have opposite charge, the mass region of the two reconstructed particles to be between 104 to 164~GeV, and have a missing transverse momentum above 50~GeV.

After selecting reconstructed mass to be between 104 to 164~GeV, eeWW becomes the dominant background. eeWW has small missing transverse momentum, as opposed to the signal, where there is a large missing transverse momentum due to $A_{H}$. 
Events before and after selection criteria are shown in Table \ref{tb:selection_eh}.  \footnote{The value in "$\sharp$ events" colomn  shows the number of events after selecting two isolated electrons with opposite charge.}   
The criteria for selecting a isolated electron is to select an electron track with an cone energy below 15~GeV, where the angle of the cone satisfies $\cos \theta \geq  0.90$. 
The criteria is optimized to maximize selection efficiency of electron coming from $e_{H}$, i.e. 84$\%$ and minimize the probability of miss identifying electrons coming from b-jet, i.e. $1.2\%$.

\begin{table}[htbp]
	\begin{center}
	\caption[event selection of analyzed processes]{\small event selection of analyzed processes}
	\begin{tabular}{rccc}
  Process                                                                                        & cross section (fb)    &   $\sharp$ of events  & \shortstack{$\sharp$ of events\\ after all cuts}  \\\hline  \hline
 $e_{H}e_{H} \rightarrow eeZ_{H}Z_{H}$                                                 &                           &  & \\ 
$ \rightarrow eeA_{H}A_{H}qqqq$			                                            	&  3.91                   & 1295          & 604\\\hline
 $ tt$                                                          &  200.7                         & 12092                &  33     \\
 $eeWW$                                                     &   1020                       &  10205               &    0  \\
  $\tau_{H}\tau_{H}$                                       &   3.32                        &  93                    &    12  \\
 $ ttZ,  ttH,  e\nu WZ,  eeZZ, WWZ, ZZZ$                    &   114.5                       &   2461                &    13 \\
\end{tabular}
\label{tb:selection_eh}
\end{center}
\end{table}

\newpage
\begin{itemize}
\item[3.] $\nu_{H}\nu_{H}$ process
\end{itemize}

The signal we have chosen for $\nu_{H}$ pair production, is a process where both $\nu_{H}$ decay to $eW_{H}$, which is further followed by $W_{H} \rightarrow WA_{H}$.
The cross section of this process is 55.7~fb at a center of mass energy of 1~TeV.
Since there are two electrons and two W bosons in the final state, we have chosen 2 electrons (one electron and one positron) and 4 jets as the signal. 
We selected events where the reconstructed W boson mass is between 60 and 100~GeV. 

\begin{table}[htbp]
		\begin{center}
		\caption[event selection of analyzed processes]{\small event selection of analyzed processes}
		\begin{tabular}{rccc}
	Process                                                                                        & cross section (fb)    &   $\sharp$ of events  & \shortstack{$\sharp$ of events\\ after all cuts}  \\\hline  \hline
	$\nu _{H}\nu _{H} \rightarrow eeW_{H}W_{H} $                                     &                            &   & \\
	$\rightarrow eeA_{H}A_{H}qqqq$  				                                     &  26.0                      & 13000  & 9132 \\\hline
	$\nu _{\tau_{H}}\nu _{\tau_{H}} , e_{H}e_{H},$                                        &   3.3 + 3.9            &   3600  & 185 \\
	$tt, ttZ, ttH, e\nu WZ, eeWW, ZZZ,WWZ $                                          &  2482                    &   1241000     & 118 \\ 
\end{tabular}   
\label{tb:selection_nh}
\end{center}
\end{table}

\subsection{Mass determination}

\begin{figure}[!htbp]%
	\begin{center}
		\subfigure[$Z_{H}Z_{H}$ process; \newline Higgs energy]{\includegraphics[width=40mm, height = 42mm]{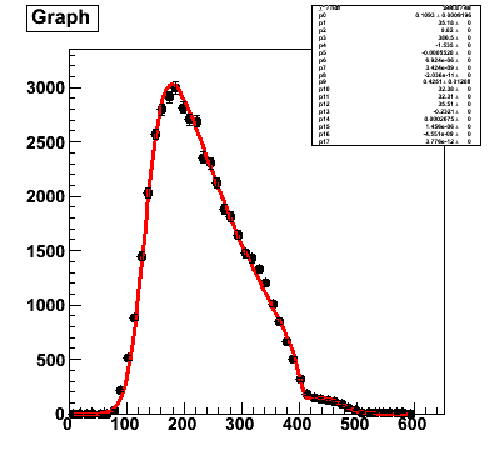}\label{fig:zh_h_e}}
		\hspace{1em}
		\subfigure[$e_{H}e_{H}$ process; \newline electron energy]{\vspace{-1em}\includegraphics[width=40mm, height = 42mm]{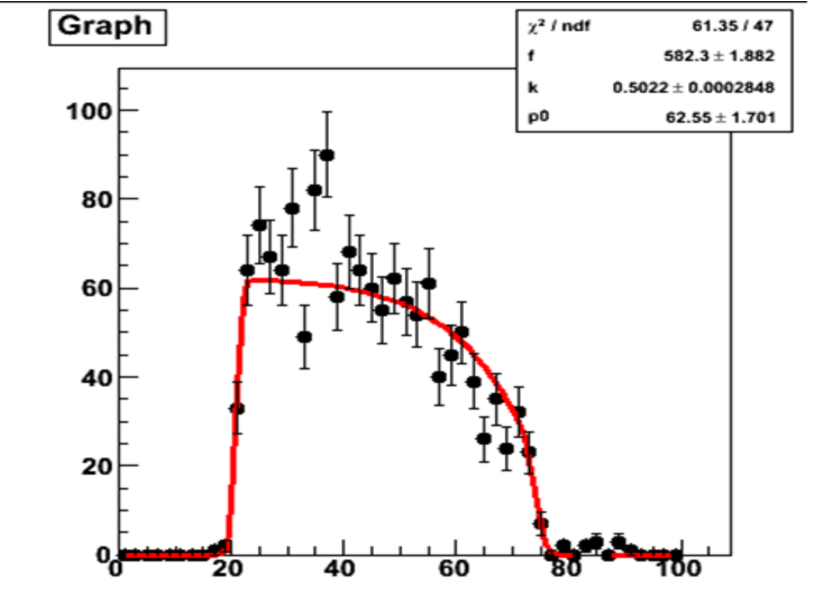}\label{fig:eh_e}}
		\hspace{1em}	
		\subfigure[$\nu_{H}\nu_{H}$ process;\newline electron energy]{\includegraphics[width=40mm, height = 42mm]{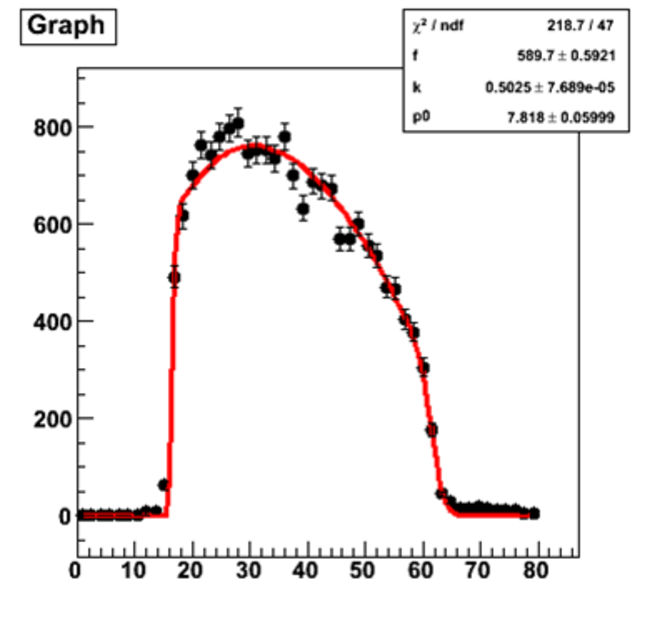}\label{fig:nuh_e}}
		\caption[energy distribution]
		{\small energy distribution}
		\label{fig:energy_dist}
\end{center}
\end{figure}

The masses of new particles can be obtained by recognizing upper and lower kinematic limits of the SMd energy distribution, from
which the masses of the new particles can be derived. 
Figure \ref{fig:energy_dist} shows the SMd energy spectrum for each process.
Figure \ref{fig:zh_h_e} shows the higgs energy distribution coming from the $Z_{H}Z_{H}$ process on top of SM background. 
Figure \ref{fig:eh_e} and figure \ref{fig:nuh_e} shows the electron energy distribution in the $e_{H}e_{H}$ and $\nu_{H}\nu_{H}$ process. Here the background is subtracted assuming the background distribution can be understood completely. \\
From $W_{H}W_{H}$ process mode we extract $W_{H}, A_{H}$ masses, from the $Z_{H}Z_{H}$ mode $Z_{H},A_{H}$ masses, from the $e_{H}e_{H} \rightarrow eZ_{H}eZ_{H}$ modes, $e_{H}, Z_{H}$ masses and from $\nu_{H}\nu_{H}$ modes $\nu_{H}, W_{H}$ masses are extracted.\\

\begin{figure}[htbp!]
	\hspace{-1.5em}	
	\begin{minipage}{0.55\hsize}
	\begin{center}
	\includegraphics[width=5cm]{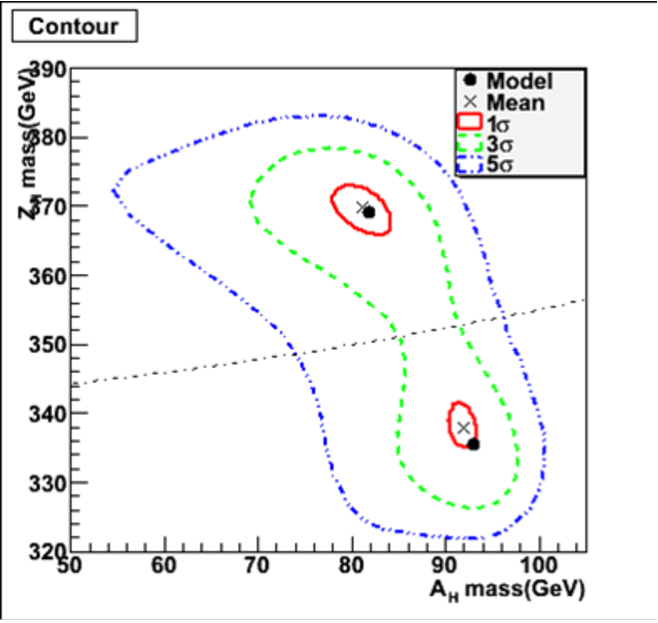}
	\end{center}
	\vskip -\lastskip \vskip -3pt
	\caption[contour plot of $Z_{H}$ and $A_{H}$ mass \newline extracted from $Z_{H}Z_{H}$ process]
 	{\small contour plot of $Z_{H}$ and $A_{H}$ mass \newline extracted from $Z_{H}Z_{H}$ process}
	\label{fig:zhzh_contour}		
	\end{minipage}
	\hspace{-3em}
	\begin{minipage}{0.55\hsize}
	\begin{center}
	\includegraphics[width=5cm]{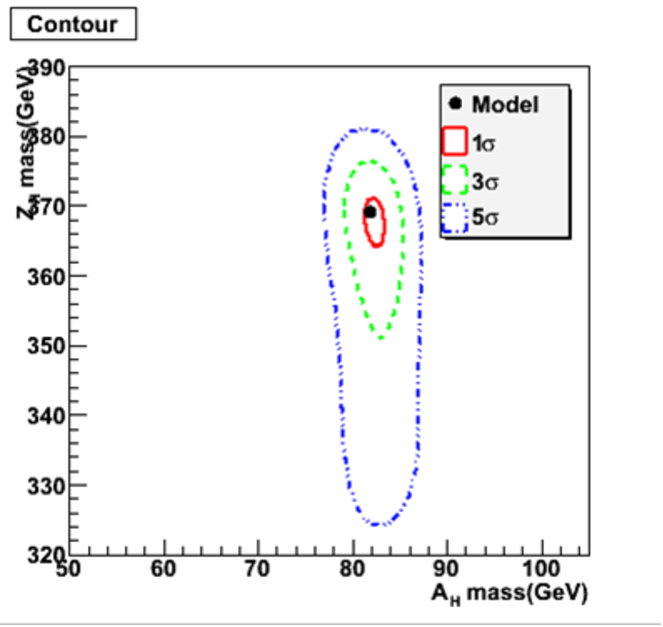}
	\end{center}
	\vskip -\lastskip \vskip -3pt
	\caption[contour plot of $Z_{H}$ and $A_{H}$ mass extracted from a simultaneous fit of $Z_{H}Z_{H}, W_{H}W_{H}$ process ]
	{\small contour plot of $Z_{H}$ and $A_{H}$ mass extracted from a simultaneous fit of $Z_{H}Z_{H}, W_{H}W_{H}$ process }
	\label{fig:zhzh_cont_simul}
\end{minipage}
\end{figure}

 Now when extracting masses from the edge of the energy distribution, we solve kinematics by solving a quadratic equation.
 This quadratic equation gives two solutions. One corresponds to a true mass solution, the other is a false solution. 
 In $e_{H}e_{H}$ $\nu_{H}\nu_{H}$ process the false mass solution is small enough that it can be excluded from previous experiments.
 However the false solution in $W_{H}W_{H}$ and $Z_{H}Z_{H}$ cannot be excluded by this method. 
 Figure \ref{fig:zhzh_contour} shows the contour plot of $A_{H}$ and $Z_{H}$ mass, which can be obtained by higgs energy distribution in the $Z_{H}Z_{H}$ process mode. 
 The two neighboring islands correspond to the two solutions. In order to select the true solution and determine $W_{H}, Z_{H}$ and $A_{H}$ masses, we performed a simultaneous fit
 using the fact that the true solution of the $A_{H}$ mass is common among both process modes, $W_{H}W_{H}$ and $Z_{H}Z_{H}$. Figure \ref{fig:zhzh_cont_simul} show the contour plot of $A_{H}$ and $Z_{H}$ mass
 after performing a simultaneous fit. A single solution can be seen.  
 The measurement accuracy of all five LHT particles masses obtained by analyzing the energy distribution in four process modes and performing a simulation fit
 is shown in Table \ref{tab:mass}.  
 
\begin{table}[!htbp]
	\begin{center}
	\caption[mass measurements accuracy]{\small mass measurement accuracy}
	\begin{tabular}{ccc}
	    particle    &  true value  &  measurement accuracy  \\ \hline \hline
		$A_{H}$ &  81.9~GeV  &  1.3$\%$   \\
		$W_{H}$ &  369~GeV & 0.20$\%$   \\
		$Z_{H}$ &  368 ~GeV & 0.56$\%$   \\
		$e_{H}$ &   410~GeV  &  0.46$\%$  \\
		$\nu _{H}$ & 400~GeV & 0.10$\%$  \\ 
		\end{tabular}
			\label{tab:mass}
			\end{center}
	\end{table}

\newpage
\subsection{Parameter extraction}

\begin{figure}[h]
	\def\captype{table}
	\begin{minipage}[t]{.40\textwidth}
    \begin{center}
    \vspace{-5em}
    \begin{tabular}{ccc}
parameter   & True value  & measurement accuracy \\ \hline \hline
		    $f$        &  580~GeV      &  $0.16~\%$    \\
        $\kappa$    &   0.5              &  $0.01~\%$    \\
\end{tabular}
\end{center}
\hspace{2em}
  \tblcaption{measurement accuracy \newline of model parameters  }
\label{tab:parameter}
\end{minipage}
  \hfill
\begin{minipage}[c]{.48\textwidth}
\hspace{3em}	
{\includegraphics[width=5cm]{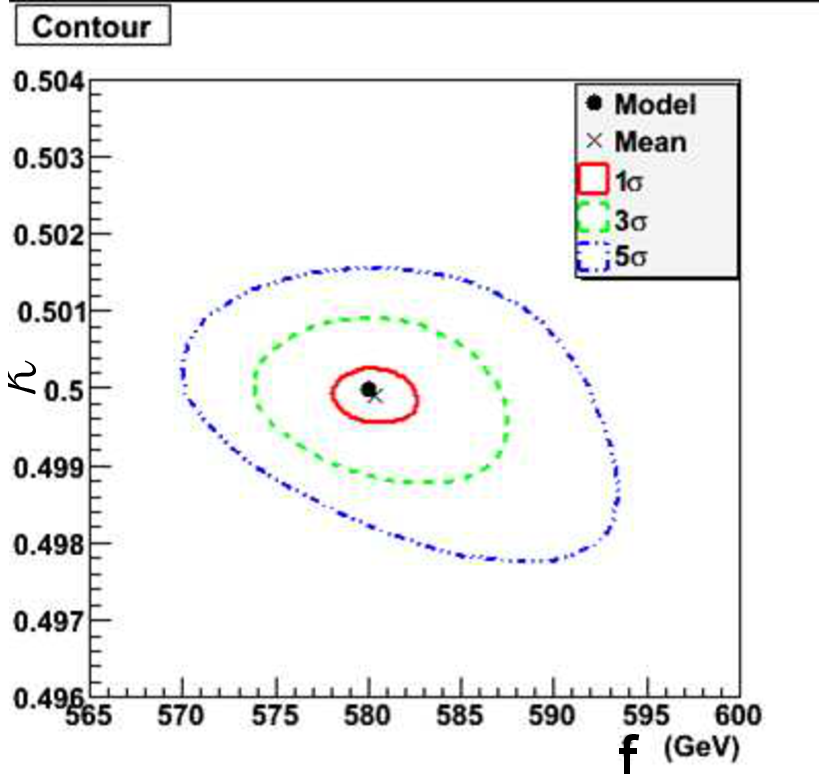}}
		\caption{contour plot corresponding to 1- and 2-$\sigma$ deviations
from the best fit point in the parameter $\kappa$ and $f$-plane}
    \label{fig:f_k}
\end{minipage}
\end{figure}

 Model parameter $f$ is extracted from the measured $A_{H}$, $Z_{H}$ and $W_{H}$ particle masses. 
 Model parameter $\kappa$ is extracted from the measured $e_{H}$ masses, which is described by $f$ and $\kappa$. 
 The contour plot of $\kappa$ and $f$ extracted in the $e_{H}e_{H}$ process mode is shown in Figure \ref{fig:f_k}.  
 The measurement accuracy of extracted model parameters are shown in Table \ref{tab:parameter}.

\subsection{Cross section measurement}

 The result of cross section measurement for all five process modes and the measurement accuracy for extracted couplings are shown in Table \ref{tab:coupling}.
 \begin{table}[htbp]
	\begin{center}
	\caption[cross section measurements]{\small cross section measurement}
	\begin{tabular}{cccc}
	process mode & extracted vertex      &cross section meas. accuracy & coupling meas. accuracy \\ \hline \hline
		$A_{H}Z_{H}$  &  $g_{ee_{H}A_{H}} $  & 7.70$\%$                          &  3.90$\%$ \\  
		$Z_{H}Z_{H}$  &  $g_{ee_{H}Z_{H}} $   &  0.859$\%$                      &  0.219$\%$\\          
		$e_{H}e_{H}$  &  $g_{e_{H}\nu_{e}W_{H}} $ &  2.72$\%$                 &  1.40$\%$\\        
		$\nu_{H}\nu_{H}$  & $g_{eW_{H}\nu_{eH}} $ &  0.949$\%$             & 0.648$\%$ \\             
		$W_{H}W_{H}$  &  $g_{e\nu_{eH}W_{H}} $ &   0.401$\%$                & 0.174$\%$ \\  
	\end{tabular}
	\label{tab:coupling}
	\end{center}
\end{table}

\newpage
\subsection{Plan. Angular distribution measurement for LHp pair production}
 We will measure production angle distribution of LHp in order to measure other couplings. 
 To derive the production angles, a quadratic equation is solved using the masses of LHp and the momenta of the SM particles, i.e. daughter of LHp with
the assumption of a back-to-back ejection of the LHp pair. The equation gives either
two solutions which contain one correct production angle or no solutions when the
discriminant of the equation is negative. The unphysical negative discriminant comes
from misreconstructing SM particle momenta or imperfect back-to-back condition of the two
LHp particle mainly due to initial state radiation.
This two dimensional production angle distribution, composed of two solutions are compared to those of templates, which are the ideal distribution of production angles with  different coupling values. We will use $\chi ^2$ (Equation \ref{eq:chi2}) to derive the $\chi^2$ minimum point and evaluate accuracy.

\section{summary}
The Littlest Higgs Model with T-parity(LHT) is one of the attractive candidates of physics beyond
the Standard Model since it solves both the little hierarchy and dark matter problems simultaneously.
 One of the important predictions of the model is the existence of new heavy gauge
bosons and heavy leptons, where they acquire mass terms through the breaking of global symmetry necessarily
imposed on the model. The determination of the masses are, hence, quite important to test
the model. 
 Once LHT masses are extracted and it is confirmed that they are consistent with the extracted model parameters; global vacuum expectation value $f$ and lepton yukawa coupling $\kappa$, which are the parameters that hold information of mass generation mechanism of the LHT model, the next important step will be to measure the couplings that appear in the process modes. 

 The couplings are described with the two model parameters. Therefore we can verify whether the measured couplings are consistent with the couplings estimated  from model parameters.  \\ We have performed Monte Carlo simulations in order to evaluate measurement accuracy of the heavy gauge bosons and heavy leptons'  masses, model parameters and couplings at the ILC. 
 The measurement accuracy of LHT particle masses are shown in Table \ref{tab:mass}.
 The measurement accuracy of LHT model parameters are shown in Table \ref{tab:parameter}, and the measurement accuracy of LHT model coupling through cross section measurement are shown in Table \ref{tab:coupling}.
 
Our plan is to evaluate the measurement accuracy of the other couplings through measurements of angular distribution.

\section{Bibliography}


\begin{footnotesize}


\end{footnotesize}


\end{document}